\documentclass[runningheads]{llncs}
\usepackage{graphicx}
\usepackage[utf8]{inputenc} 
\usepackage[T1]{fontenc}    
\usepackage{hyperref}       
\usepackage{url}            
\usepackage{booktabs}       
\usepackage{amsfonts}       
\usepackage{nicefrac}       
\usepackage{microtype}      
\usepackage{xcolor}         
\usepackage{multirow}

\begin{document}
\title{NeuralMind-UNICAMP at 2022 TREC NeuCLIR:\\
Large Boring Rerankers for Cross-lingual Retrieval
}
%
%
\author{Vitor Jeronymo \and
Roberto Lotufo \and
Rodrigo Nogueira}
%
%
\institute{NeuralMind, Brazil\\
UNICAMP, Brazil}
\maketitle              
\begin{abstract}

This paper reports on a study of cross-lingual information retrieval (CLIR) using the mT5-XXL reranker on the NeuCLIR track of TREC 2022. Perhaps the biggest contribution of this study is the finding that despite the mT5 model being fine-tuned only on query-document pairs of the same language it proved to be viable for CLIR tasks, where query-document pairs are in different languages, even in the presence of suboptimal first-stage retrieval performance. The results of the study show outstanding performance across all tasks and languages, leading to a high number of winning positions. Finally, this study provides valuable insights into the use of mT5 in CLIR tasks and highlights its potential as a viable solution. For reproduction refer to \url{https://github.com/unicamp-dl/NeuCLIR22-mT5}
  
\keywords{CLIR \and reranker \and mT5 \and IR \and NeuCLIR.}
\end{abstract}
\section{Introduction}



The increasing global interconnectedness of society and the growing need for access to information in multiple languages have brought the importance of cross-lingual information retrieval (CLIR). As businesses expand their operations across borders and individuals become more multilingual, the requirement for effective retrieval of information in multiple languages has become increasingly compelling.

The TREC conference has recently launched the NeuCLIR dataset challenge, which focuses on CLIR. The corpus includes 4.5 million Russian, 3 million Chinese, and 2 million Persian documents obtained from the Common Crawl service between August 1, 2016 and July 31, 2021. The dataset also features machine translations of these documents into English, as well as 114 English queries, along with their machine and human translations into the three languages.

The dataset consists of three tasks: Ad Hoc CLIR, Reranking CLIR, and Monolingual Retrieval. Ad Hoc CLIR represents the standard retrieval pipeline for IR datasets, while Reranking CLIR involves reranking the provided query-document pairs. Monolingual Retrieval is similar to Ad Hoc, but uses human-translated topics.

In order to address the challenges posed by the NeuCLIR dataset, we developed a pipeline that utilizes machine translation to align the language of the queries with that of the passages. This is followed by a multi-stage retrieval pipeline that includes BM25~\cite{Robertson1994OkapiAT} as the first stage retriever and a large multilingual T5 model as the second stage, proven to be an important component in top-performing systems in various monolingual and multilingual IR benchmarks~\cite{qiao2020pash,Qiao2022PASHAT,Huang2022YorkUA,gao2022unsupervised,zhang2022hlatr,jeronymo2022boring,jeronymo2022mrobust04,bonifacio2021mmarco}. Our method takes advantage of the observation that increasing the number of parameters in a model generally leads to improved retrieval effectiveness~\cite{rosa2022no}.


\section{Related Work}

Information Retrieval (IR) has a long and robust history in research, with numerous publicly accessible datasets developed to support its advancement. Some of the most well-known and widely-used ones are MS-MARCO~\cite{MS_MARCO_v3}, TREC~\cite{Voorhees_TREC2004_robust,zhang2020rapidly,craswell2020overview}, Common Crawl~\cite{ccrawl}, and ClueWeb22~\cite{overwijk2022clueweb22}. In response to the challenges of IR, various models and methods have been proposed, utilizing both classic ones such as Vector Space Model (VSM)~\cite{vsm}, Latent Semantic Indexing (LSI)~\cite{deerwester1990indexing} and BM25, as well as more modern  transformer-based models, such as RoBERTa~\cite{liu2019roberta}, BERT~\cite{devlin2018bert}, and T5~\cite{raffel2020exploring}. The latter ones have shown effectiveness in handling multiple tasks, such as text classification~\cite{tezgider2022text,zhang2021fast,li2021act}, named entity recognition~\cite{yan2019tener,arkhipov2019tuning,lothritz2020evaluating}, and IR~\cite{lin2021pretrained}.

Regarding the branch of multilingual and CLIR, it is crucial to have access to appropriate datasets that can be used for both development and evaluation of models. In recent years, several datasets that support research in this area have been made publicly available, such as Fire~\cite{mitra2008overview,majumder2013overview}, MLQA~\cite{lewis2019mlqa}, NTCIR~\cite{sakai2021evaluating}, Mr. Tydi~\cite{zhang2021mr}, and HC4~\cite{HC4}.

However, traditional lexical algorithms such as BM25~\cite{Robertson1994OkapiAT} face challenges in providing efficient multilingual capabilities, as they rely on exact word matches between the query and the passage for ranking relevance. To overcome this, one solution is to use automatic translators, such as GoogleTranslate, BingTranslate, Helsinki~\cite{tiedemann2020opus}, or Facebook's WMT19~\cite{ng2019facebook}, to transform either the query or the passage into a common language.

As seen in monolingual, transformer-based models have also shown to be effective in handling multilingual and cross-lingual tasks, and there are various approaches to leverage their potential. One strategy is to fine-tune a pre-trained model on a specific task and language, using either zero-shot or few-shot learning techniques. Another approach is multilingual learning, where a pre-trained model is fine-tuned on a multilingual corpus. A third strategy is cross-lingual transfer learning, which involves fine-tuning a pre-trained model on a specific task for multiple languages. 

In addition to these fine-tuning approaches, there are also models that are pre-trained on a massive multilingual corpus, such as mT5~\cite{xue2020mt5}, XLM-R~\cite{conneau2019unsupervised}, and mBERT~\cite{devlin2018bert}, which are specifically designed for multilingual scenarios. Increasing the number of parameters in a model also tends to enhance retrieval effectiveness. In light of the aforementioned considerations, this study proposes the implementation of a reranking mechanism for the specified task, and investigates the effectiveness of evaluating the mT5-XXL model (with 13B parameters) in a zero-shot manner.

\section{Methodology}

In this section, we outline the models used for the first-stage retrieval and reranking stages of our CLIR system. We explain how the reranker was trained for the task, and how we arrived at the optimal configuration for our final submission.

In order to perform the first stage retrieval, we utilized the runs provided by the organizers of the competition, as well as the BM25~\cite{Robertson1994OkapiAT} retrieval method implemented in the Pyserini framework~\cite{lin2021pyserini} and the SPLADE~\cite{formal2021splade}. To accommodate for the cross-lingual nature of the corpus and queries, we employed various automatic machine translation tools such as Google Translate, Microsoft Bing Translator, Facebook, Huawei, Caiyun, and Youdao to translate the English queries into the target languages of Persian, Chinese, and Russian. These runs, which utilized automatic machine translation, are classified as Ad Hoc Cross-Language Information Retrieval.

We utilized the multilingual variant of the T5 model, mT5-XXL to rerank the runs provided by the first-stage retrievers. We fine-tuned it in the same manner as Bonifacio et al.~\cite{bonifacio2021mmarco}, on the mMARCO dataset utilizing a batch size of 128 and a maximum sequence length of 512 tokens. The model was fine-tuned for 100,000 iterations with a constant learning rate of 0.001, and all layers incorporated a dropout of 0.1, taking around 114 hours on a TPU v3-8. It is noteworthy that the standard training procedure in mMarco is multilingual in nature, with query and document samples being in different languages, yet each sample pair is in the same language.

The HC4~\cite{lawrie2022hc4} dataset was employed as a validation set for the selection of the most optimal translators and first-stage retrievers  due to the shared language coverage between it and NeuCLIR, as well as the existence of overlapping annotated query-document pairs between the two datasets. The RRF and SPLADE first-stage runs were provided by the NLE and h2loo teams, however, at the time of submission, the NLE team did not have a SPLADE model available for Chinese.

The selection of the optimal query translator for each language was performed using the nDCG@20 and R@1000 metrics. The results presented in Table ~\ref{tab:first_stage_HC4} indicate that the Bing translator was the most effective for Persian and Russian, while the Youdao translator was deemed the best option for Chinese. 

\begin{table}[]
\centering
\begin{tabular}{cllrrrrr}
\hline
 &  &  & \multicolumn{1}{l}{NDCG@20} & \multicolumn{1}{l}{MAP} & \multicolumn{1}{l}{RBP} & \multicolumn{1}{l}{R@100} & \multicolumn{1}{l}{R@1000} \\ \hline
\multirow{8}{*}{fa}  & \multirow{2}{*}{Bing}              & desc \& title & 0.398 & 0.287 & 0.281 & 0.646 & 0.828 \\
                     &                                    & title         & 0.345 & 0.261 & 0.255 & 0.577 & 0.772 \\ \cline{2-8} 
                     & \multirow{2}{*}{Facebook}          & desc \& title & 0.381 & 0.274 & 0.279 & 0.622 & 0.840 \\
                     &                                    & title         & 0.297 & 0.224 & 0.232 & 0.541 & 0.739 \\ \cline{2-8} 
                     & \multirow{2}{*}{Huawei}            & desc \& title & 0.336 & 0.243 & 0.242 & 0.562 & 0.767 \\
                     &                                    & title         & 0.212 & 0.155 & 0.154 & 0.399 & 0.621 \\ \cline{2-8} 
                     & \multirow{2}{*}{Human translation} & desc \& title & 0.449 & 0.329 & 0.316 & 0.689 & 0.865 \\
                     &                                    & title         & 0.385 & 0.288 & 0.274 & 0.636 & 0.822 \\ \hline
\multirow{6}{*}{ru}  & \multirow{2}{*}{Bing}              & desc \& title & 0.308 & 0.232 & 0.253 & 0.536 & 0.766 \\
                     &                                    & title         & 0.315 & 0.231 & 0.266 & 0.492 & 0.712 \\ \cline{2-8} 
                     & \multirow{2}{*}{Huawei}            & desc \& title & 0.317 & 0.240 & 0.268 & 0.526 & 0.768 \\
                     &                                    & title         & 0.290 & 0.204 & 0.249 & 0.475 & 0.728 \\ \cline{2-8} 
                     & \multirow{2}{*}{Human translation} & desc \& title & 0.298 & 0.227 & 0.251 & 0.529 & 0.772 \\
                     &                                    & title         & 0.295 & 0.219 & 0.248 & 0.487 & 0.718 \\ \hline
\multirow{10}{*}{zh} & \multirow{2}{*}{Bing}              & desc \& title & 0.274 & 0.192 & 0.205 & 0.462 & 0.674 \\
                     &                                    & title         & 0.241 & 0.179 & 0.186 & 0.466 & 0.659 \\ \cline{2-8} 
                     & \multirow{2}{*}{Caiyun}            & desc \& title & 0.276 & 0.195 & 0.198 & 0.500 & 0.687 \\
                     &                                    & title         & 0.263 & 0.187 & 0.190 & 0.442 & 0.635 \\ \cline{2-8} 
                     & \multirow{2}{*}{Huawei}            & desc \& title & 0.273 & 0.185 & 0.198 & 0.468 & 0.663 \\
                     &                                    & title         & 0.237 & 0.163 & 0.173 & 0.455 & 0.640 \\ \cline{2-8} 
                     & \multirow{2}{*}{Human translation} & desc \& title & 0.260 & 0.191 & 0.187 & 0.484 & 0.710 \\
                     &                                    & title         & 0.253 & 0.180 & 0.178 & 0.490 & 0.696 \\ \cline{2-8} 
                     & \multirow{2}{*}{Youdao}            & desc \& title & 0.268 & 0.197 & 0.197 & 0.473 & 0.673 \\
                     &                                    & title         & 0.235 & 0.171 & 0.172 & 0.463 & 0.645 \\ \hline
\end{tabular}
\vspace{0.2cm}
\caption{Comparison of different machine translators on the test set of the HC4 dataset.}
\label{tab:first_stage_HC4}
\end{table}

After conducting multiple experiments, as illustrated in Table ~\ref{tab:hc4}, we sought to determine the optimal configuration for the reranker with regards to the first-stage retriever, query translator, and type of query (title, description, or both) to be input into the model. The most effective combination was found to be:

\begin{itemize}

  \item {Persian and Russian:} Bing as query translator; SPLADE as a first-stage retriever, followed by an mT5 reranker that uses the ``description'' field of queries;
  
  \item {Chinese:} Youdao as query translator; BM25 RRF as a first-stage retriever; mT5 as a reranker that uses the concatenation of fields ``title'' and ``description'' as queries. For the reranker, we use the original query in English.

\end{itemize}

All first-stage retrievers use the concatenation of fields ``title'' and ``description'' as queries and the corpora were in their respective original languages.

\begin{table}[]
\centering
\begin{tabular}{clccccc}
\hline
\multicolumn{1}{l}{Language} & Run                           & NDCG@20 & MAP    & RBP    & R@100  & R@1000 \\ \hline
\multirow{6}{*}{fa}          & BM25 (Bing translated)        & 0.3984  & 0.2873 & 0.2812 & 0.6460 & 0.8275 \\
                             & +description+title-mT5-en     & 0.5558  & 0.4584 & 0.4143 & 0.7934 & 0.8275 \\
                             & +description+title-mT5-bing   & 0.5639  & 0.4807 & 0.4332 & 0.7900 & 0.8275 \\
                             & +description-mT5-bing         & 0.5813  & 0.4843 & 0.4353 & 0.7859 & 0.8275 \\ \cline{2-7} 
                             & RRF HT                   & 0.4104  & 0.3172 & 0.3066 & 0.6878 & 0.9044 \\
                             & +description+title-mT5-HT   & 0.6000  & 0.5085 & 0.4469 & 0.8531 & 0.9044 \\ \hline
\multirow{5}{*}{ru}          & BM25 (Bing translated)        & 0.3078  & 0.2317 & 0.2527 & 0.5357 & 0.7664 \\
                             & +description+title-mT5-en     & 0.4369  & 0.3611 & 0.3670 & 0.6746 & 0.7664 \\
                             & +description-mT5-bing         & 0.4513  & 0.3820 & 0.3829 & 0.7015 & 0.7664 \\ \cline{2-7} 
                             & RRF HT                   & 0.3262  & 0.2624 & 0.2842 & 0.5873 & 0.8206 \\
                             & +description+title-mT5-HT   & 0.4535  & 0.3931 & 0.3783 & 0.7005 & 0.8206 \\ \hline
\multirow{8}{*}{zh}          & BM25 (Bing translated)        & 0.2744  & 0.1916 & 0.2047 & 0.4623 & 0.6737 \\
                             & +description+title-mT5-en     & 0.5539  & 0.4352 & 0.4154 & 0.6589 & 0.6737 \\ \cline{2-7} 
                             & BM25 (Youdao translated)      & 0.2679  & 0.1972 & 0.1973 & 0.4734 & 0.6732 \\
                             & +description+title-mT5-youdao & 0.5516  & 0.4405 & 0.4188 & 0.6649 & 0.6732 \\
                             & +description+title-mT5-en     & 0.5533  & 0.4382 & 0.4144 & 0.6610 & 0.6732 \\
                             & +description+title-mT5-mt     & 0.4667  & 0.3689 & 0.3548 & 0.6330 & 0.6732 \\ \cline{2-7} 
                             & RRF HT                   & 0.2792  & 0.2127 & 0.2127 & 0.4940 & 0.7492 \\
                             & +description+title-mT5-HT & 0.5975  & 0.4784 & 0.4410 & 0.7356 & 0.7492 \\ \hline
\end{tabular}
\vspace{0.2cm}
\caption{HC4 results for first-stage retrieval (BM25, SPLADE or RFF) followed by an mt5-XXL reranker. ``HT'' stands for human-translated queries.}
\label{tab:hc4}
\end{table}

\section{Results}

The NeuCLIR challenge submissions were constructed using SPLADE and BM25 RRF runs provided by the NLE and h2loo teams, similar to the previous HC4 experiments. The best configuration found in the HC4 experiments was applied to the NeuCLIR submissions, as illustrated in Table~\ref{tab:neuclir}. However, contrary to the expectations, the results showed that using SPLADE as the first-stage retriever resulted in the worst performance. The best run for Persian was achieved by reranking the organizer's run, for Russian, the best run utilized BM25 RRF, and for Chinese, the best nDCG@20 was achieved with RRF and human-translated queries.

It is important to note that subsequent to conducting our experiments, we discovered a bug in some of the BM25 RRF runs. This resulted in a significant decrease in the performance of the Chinese BM25 RRF run, as evident from its low R@1000 score of 0.4539. Given the impact of this bug, it is likely that our runs using BM25 RRF as first-stage retrievers would have performed better if this issue had not arisen. Unfortunately, at the current time, we are unable to provide updated metrics as we do not have access to the NeuCLIR annotated query-document relevance data (qrels).

\begin{table}[]
\begin{tabular}{l|ccc|ccc|ccc}
\hline
\multicolumn{1}{c|}{\multirow{2}{*}{1st-stage}} & \multicolumn{3}{c|}{fa}  & \multicolumn{3}{c|}{ru}  & \multicolumn{3}{c}{zh}                                             \\ \cline{2-10} 
\multicolumn{1}{c|}{} & nDCG@20 & mAP    & R@1000 & nDCG@20 & mAP    & R@1000 & nDCG@20 & mAP    & R@1000 \\ \hline
SPLADE                                          & 0.5356 & 0.3971 & 0.8069 & 0.5524 & 0.4167 & 0.7642 & \multicolumn{1}{l}{} & \multicolumn{1}{l}{} & \multicolumn{1}{l}{} \\
Organizer's run       & 0.5881  & 0.4350 & 0.8292 & 0.5483  & 0.4216 & 0.7744 & 0.5165  & 0.4039 & 0.7814 \\
HT RRF                & 0.5619  & 0.4215 & 0.8329 & 0.5673  & 0.4387 & 0.7612 & 0.4999  & 0.3837 & 0.7022 \\
RRF                   & 0.5447  & 0.4042 & 0.7820 & 0.5630  & 0.4340 & 0.8036 & 0.3689  & 0.2177 & 0.4539 \\
2nd best              & 0.545   & 0.404  & 0.782  & 0.565   & 0.473  & 0.898  & 0.484   & 0.360  & 0.750  \\ \hline
TREC Median           & 0.3200  & 0.1983 & 0.8195 & 0.3725  & 0.2578 & 0.7590 & 0.2811  & 0.1854 & 0.7269 \\
TREC Max              & 0.7340  & 0.5963 & 0.9556 & 0.7053  & 0.5694 & 0.9445 & 0.6847  & 0.5600 & 0.9136 \\ \hline
\end{tabular}
\caption{Main results on NeuCLIR.}
\label{tab:neuclir}
\end{table}

\begin{figure}[h]
  \centering
  \includegraphics[width=13cm]{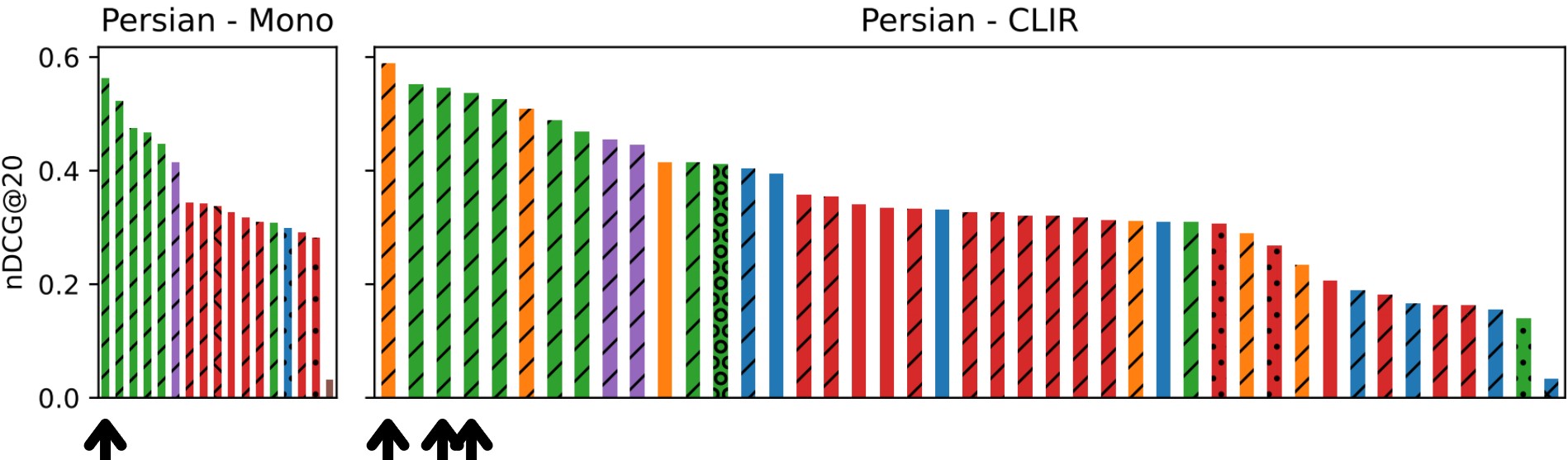}
  \caption{Persian Leaderboard. The arrows illustrated in black denote our submissions.}
  \label{fig:neuclir_persian}
\end{figure}

\begin{figure}[h]
  \centering
  \includegraphics[width=13cm]{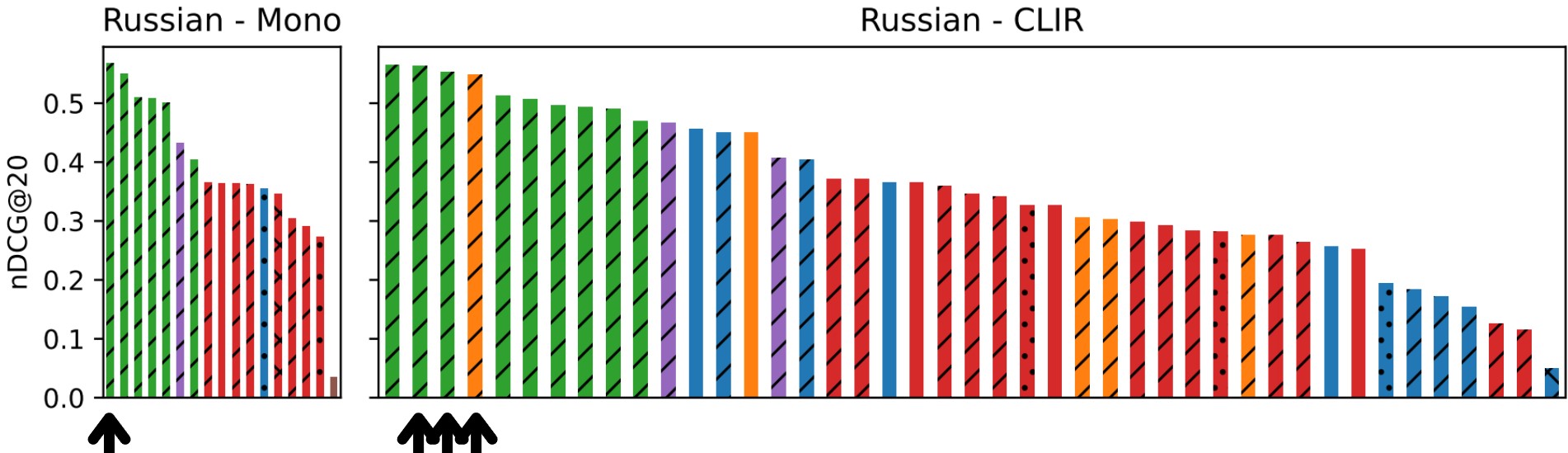}
  \caption{Russian Leaderboard. The arrows illustrated in black denote our submissions.}
  \label{fig:neuclir_russian}
\end{figure}

\begin{figure}[h]
  \centering
  \includegraphics[width=13cm]{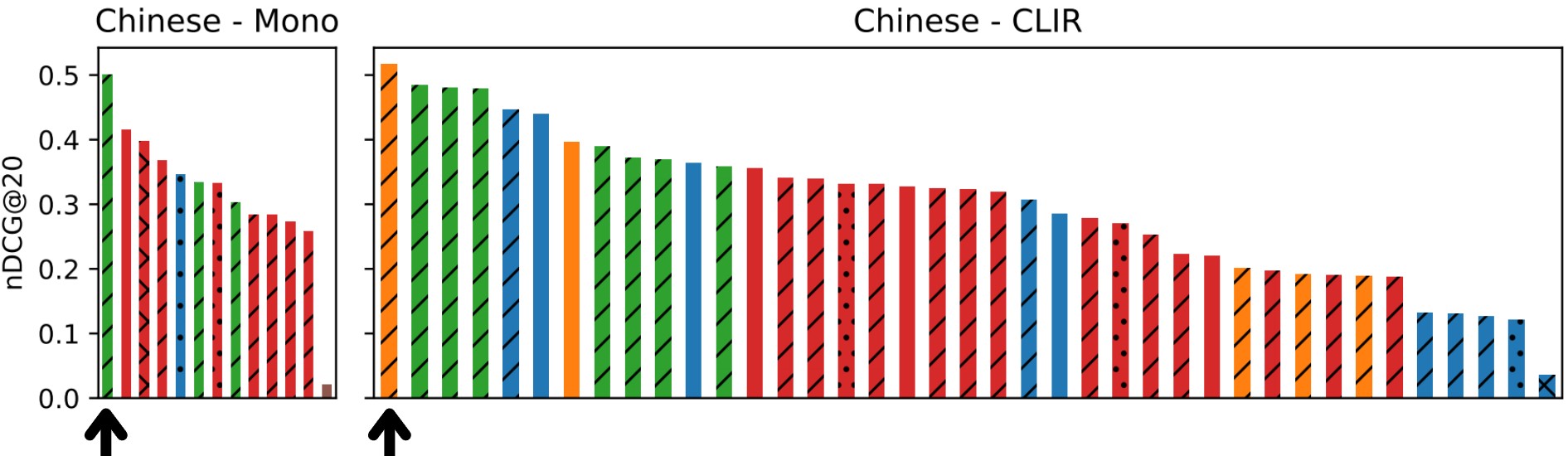}
  \caption{Chinese Leaderboard. The arrows illustrated in black denote our submissions.}
  \label{fig:neuclir_chinese}
\end{figure}

\section{Conclusion}

Our evaluation results showed that most of our runs performed similarly regardless of the first-stage retriever used. However, SPLADE performed poorly, with a lower recall compared to even the faulty BM25 RRF runs. Additionally, our mT5 reranker demonstrated its robustness even with the faulty RRF runs as first-stage retrievers. The reranker was able to surface relevant documents to the top of the list, making our submissions still competitive in the competition, with some even surpassing the results of the organizers' runs. This highlights the robustness of the reranker, even in the presence of suboptimal first-stage retrieval performance.

The evaluation of the model on a CLIR task marked a noteworthy achievement in our research. Despite being fine-tuned on the mMarco dataset in a multilingual context, rather than a cross-lingual one, mT5 demonstrated its viability as a reranker for CLIR scenarios. This finding highlights the versatility of mT5 in adapting to different tasks and languages.

Overall, our submissions exhibited exceptional performance across all languages and achieved top ranking positions in the majority of tasks, as demonstrated by a substantial margin of superiority when compared to other teams' submissions. This is illustrated by the black arrows in Figures \ref{fig:neuclir_persian}, \ref{fig:neuclir_russian}, and \ref{fig:neuclir_chinese}, which pertain to the Persian, Russian, and Chinese languages, respectively.

%
%
%
\bibliographystyle{splncs04}
\bibliography{main}
\end{document}